# Frequency Regulation in Islanded Microgrid Using Demand Response

Alireza Eshraghi[1], Mahdi Motalleb[2], Ehsan Reihani[3], Reza Ghorbani[2]
[1]University of California, Riverside, Email: seshr001@ucr.edu
[2]University of Hawaii at Manoa, Emails: {motalleb, rezag}@hawaii.edu
[3]California State University, Bakersfield, Email: ereihani@csub.edu

*Abstract*—Introducing more Distributed Generation (DG) into power grid infrastructure drives more attention to understand how large scale DG affects grid operation. Islanding is an important concern in this area. Islanding refers to the condition that DGs within a microgrid continue energizing while the microgrid has been disconnected from the main grid. Considering the adverse effects of Islanding, it should be detected and managed in a proper way. After Islanding has been detected, even though the first option is tripping all inverter-based DG unit in the system, the system has this option to work as a stand-alone microgrid. For achieving this goal, the frequency of microgrid should be regulated. This paper proposes an islanding detection method based on current detection in a parallel arm with the fuse arm of the microgrid. After islanding detection, this paper presents an effective implementation Demand Response (DR) to regulate the frequency in islanded microgrid as an Ancillary Service (AS) considering the transient constraints of frequency in inverter-based generations.

## I. Introduction

Maximizing the security of microgrid is desirable from a system designer viewpoint. In such a system all components work in the safe operation range and they have the ability to continue operation during system failures until the operator restore the system. Islanding detection is one of the critical issues in this context. During the islanding, the system should operate within an acceptable reliability and also be ready to connect to the main grid at the appropriate time. With increasing penetration of renewable energy in the microgrid, keeping the safe and sound operation of the microgrid has become more challenging. When a photovoltaic (PV) system or other Distributed Generation (DG) continue to power a part of the grid while that part has been isolated from the main electric utility islanding will occur [1].

Finding efficient methods in order to avoid islanding or at least control the islanded microgrid is critical to have a more reliable power system. Personal safety hazards, equipment damage, grid protection interference, and power quality issues are some of the adverse effects of islanding. IEEE-1547, UL1741, and IEC-62116 are some of the standards that are used to provide guidance when it comes to implementing DGs into a pre-existing network infrastructure [2, 3, 4].

Islanding detection methods can be divided into three different categories: passive, active, and communication based methods. In the passive methods a local parameter is compared with a predetermined threshold. A considerable difference shows that microgrid has been islanded. Having large None Detection Zone (NDZ) area makes these methods insufficient for the anti- islanding protection, although they have the advantages of being low cost and easily implemented. On the other hand, active methods are based on injecting a small disturbance into the system and monitoring the response of the system to decide whether microgrid is islanded or not. The injected disturbance will be corrected by the grid if the microgrid is still connected to it (?). However, system parameters will be changed considerably by the injection if the system is islanded. Power quality issues, being ineffective when the system has more than one DG are some disadvantage of this method even though it has the advantage of smaller NDZ area in comparison with passive ones [1, 5].

Although the first option is to trip the DG units after islanding detection, the second option is managing the microgrid to work perfectly and stably as a stand-alone system. After the islanding has occurred, the microgrid should operate stably using all available resources including DGs and also ancillary services (AS) which are contracted to stand by the microgrid to manage power delivery for balancing microgrid condition on short notice. AS refers to a variety of operations including regulation up, regulation down, spinning reserve, non-spinning reserve and reactive power. Purchasing for being available in the emergency situation is the difference between AS and other type of power products. Frequency regulation is one of the AS services that can be done in the islanding situation when the microgrid is disconnected from the main grid in order to maintain the system balance. For AS integration and managing, the islanding situation the amount that should be purchased, prioritizing the resources to be used, and defining who should pay for these services should be planned [6, 7, 8, 9, 21].

Frequency regulation can be attained using Demand Response (DR) both as a load and as generation. Aggregate of DR resources can be used to absorb the excess power generation from renewable resources when the generators produce at their minimum generation level. In [10, 11, 12, 22], DR are managed to participate in market in order to provide AS such as contingency reserve. A market model is proposed and profits of Generation Companies (GENCOs) and Demand Response Aggregators (DRAs) are determined after participation in DR market. In [13], impact of DR delay on the frequency of power system is investigated and fuzzy-PI based controller is used to remove the delay. DR is also used for primary frequency regulation in microgrid when there is an intermittent generation from wind turbines [14]. Commercial heating, ventilation, and air conditioning (HVAC) are large loads which can have a high effect on frequency and thus are a potential candidate for frequency regulation [15]. Different control strategies for providing frequency regulation from commercial HVAC systems and components are investigated and discussed. In case of islanding, wireless sensor networks can be utilized to detect islanding [16, 17] and operate available DR resources to regulate the frequency in island mode.

The focus of this paper is to propose a new islanding detection method which can overcome typical challenges associated with other islanding detection methods. After islanding detected, this paper describes an effective implementation of DR as an AS instead of compulsory provision to handle the islanded microgrid. The paper defines this compulsory provision as Islanding Reserve (IR) which should be provided by GENCOs for maintaining the microgrid stable in the situation of islanding. DR can be used to provide some sort of AS to the system by providing some voluntary load shedding in the microgrid. Through this way generating units can be used for stabilizing the microgrid in the situation of islanding instead of using IR. Additionally, the failure probability of demand side to provide power in the critical situation is smaller than generating units since this service would be provided by combination of a large number of small loads which are less likely to fail at the same time. So it is a more reliable solution for managing the islanding condition [18].

The rest of this paper is organized as follows. Section II describes the proposed islanding detection method. Section III explains how the proposed method manage the microgrid in the islanding situation as a standalone system. In Section IV the simulation results are presented and analyzed. Finally, the conclusion is drawn in Section V.

## II. ISLANDING DETECTION

Figure 1 shows a schematic diagram of the proposed method for islanding detection. In this method an impedance arm is placed in parallel with the Main Switch (MS). MS is the connection point of microgrid to the main grid and islanding is occurred when this switch is disconnected. In the normal grid connected operation, power exchange between microgrid and main grid is done through the MS since the Parallel Impedance Arm (PIA) has been short circuit by MS. When the system is islanded the whole current passes through a PIA which is equipped with the current sensor and an operable switch. A coordinator monitors the current sensor output and control the switch in PIA. When PIA's current is detected (or it is more than a threshold value), then islanding is detected by the system. In order to stop energizing the microgrid completely, coordinator turns off the operable switch right after detecting the islanding situation. After islanding detected by the system, two scenarios might be initiated. The first one is tripping the DGs in the system. This method is currently in use with other anti-islanding methods. In this method after islanding has been detected the coordinator sends tripping command to all DGs' inverters using communication methods such as ZigBee, RFID, or power line [19]. The other option is managing the microgrid to work as a standalone system perfectly and stably after islanding has been detected [20]. For achieving this, the frequency of the microgrid should be regulated which is the topic of the next section.

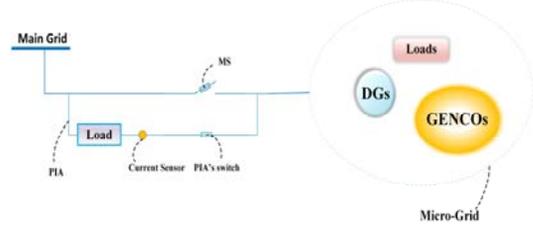

*Figure 1. Micro Grid equipped with the proposed method*

## III. Frequency Regulation in the Island

### A. Frequency Control Method

Frequency change is a result of an imbalance between load and generation. In order to control the frequency, the generator output is controlled by frequency controller to remove the imbalance. For this goal, the mechanical torque should be converted to electrical energy completely. In islanded microgrids with high penetration of renewable energy generation, as a result of decreasing the actual mass of physical rotating, the moment of inertia is reduced. This reduction makes the entire system more sensitive to imbalances and frequency droop. Equation (1) shows the dynamic relationship between mechanical power ($P_m$) and electrical power ($P_e$) using swing equation where $\delta$ is the electrical power angle, $\omega_s$ is the electrical angular velocity, and $H$ is per unit inertia constant (second). Since the purpose of the model is for a small microgrid, it can be modeled with a single swing equation.

$$\frac{2H}{\omega_S}\frac{d^2\delta}{dt^2} = P_m - P_e \quad (1)$$

Load changes ($\Delta P_L(s)$) in islanded microgrid effects the changes in output frequency ($\Delta\Omega(s)$). This relationship is expressed in equation (2) where $\tau_g$ and $\tau_T$ are the governor and turbine time constants, $D$ is load change in percent over frequency change in percent, and $R$ is speed regulation in percent.

$$\frac{\Delta\Omega(s)}{-\Delta P_L(s)} = \frac{(1+\tau_g s)(1+\tau_T s)}{(2Hs+D)(1+\tau_g s)(1+\tau_T s)+1/R} \quad (2)$$

With neglecting the losses in the microgrid, generation reduction can be supposed to be equal to load increase with an equivalent swing equation. Therefore, equation (2) can analyze the transient and dynamic behaviors of islanded microgrid. Equation (3) shows deviation of the steady state frequency:

$$\Delta \omega_{ss} = (-\Delta P_L) \frac{1}{D + 1/R} \quad (3)$$

DGs work according to interconnection standard IEEE 1547 [2]. Inverter-based DG units are designed to trip after abnormal operating frequency detection out based on this standard.

### B. Market; competition between aggregator

The profit of all participating parties is determined by clearing the market. In this market, IR can be purchased directly from aggregators using DR in the microgrid instead of utility. The proposed market framework has three main parts: first we calculate the cost which the generator(s) of the islanded grid pay to regulate the frequency in the absence of DR sources through increasing the generation (subsection 1 below). Second Demand Response Aggregators (DRAs) make a market to submit their bids and compete to each other to shed the loads in order to regulate the frequency (subsection 2 below). And finally, after clearing the market, profits of GENCO and DRAs are calculated based on their participation in the market (subsection 3 below).

*1- IR cost curve for island's GENCO using OPF*

Island's *GENCO* provides the cost curves during the islanding. After islanding detected, the first step for the algorithm is to calculate the transient of frequency and new inertia. Based on IEEE-1547 standard [2], when the overshoot is less than 3 Hz, the fault needs to be cleared in less than 300s and for overshoot greater than 3Hz the time is 160ms. Unsuccessful clearance in this time frame leads to trip off for the inverter-based generations in the microgrid. In the case when the overshoot of frequency is greater than 3 Hz, Island's *GENCO* starts to recover the frequency through its available reserves to at least 57 Hz within 160ms since DR is not fast enough to do the recovery in this short time. This amount of power is considered as emergency reserved. After the frequency reached to 57 Hz, the aggregators can compete with each other to sell DR to the *GENCO*. Clearly if the frequency is above 57Hz after islanding, the IR cost includes only the non-emergency reserve which can be supplied either by the utility or aggregator through DR.

For frequency regulation where DR is responsible the non-emergency (57-60 Hz) zone is split into *M* sections which can be equal or unequal. Required power ($\Delta P$) for increasing the frequency in each section ($\Delta f$) by utility is calculated using equation (2). It should be considered that inertial constant (*H*) changes with power increasing in each section. According to IEEE-1547 standard [2], when the grid frequency is in the range of 57-58.5 Hz, the required power should be supplied within 300s. Otherwise the inverter-based generations (PVs) will trip off. Because of that, the non-emergency range can be divided in two sections: 1) 57-58.5 Hz where faster DR is needed, and 2) 58.5-60 Hz where the slower DR starts. The point where optimal grid frequency condition archives is the last point of the utility cost curve. Figure 2 explains the overall concept of the method which how *GENCO* and DRAs work together to regulate the frequency after islanding. Figure 3 shows the flowchart for IR cost curve (subsection 1).

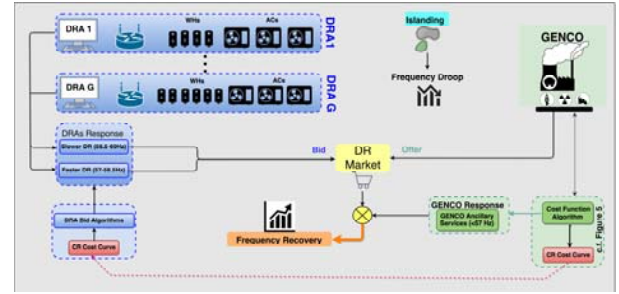

Figure 2. Demand response market implementation for open trading of ancillary services in the island

*2-Forming DR market for the aggregators in the island*

For utilizing DR as a source of AS in the islanded microgrid to provide counter bids, a market needs to be constructed. It is considered that *G* aggregators participate in the DR market where each of them holding contracts with undefined number of costumers. The bid prices are $F_1^{DR}, F_2^{DR}, ..., F_G^{DR}$. Each aggregator allocates its resources based on the speed of DR response. DR categorize as faster DR for 57-58.5 Hz and slower DR for 58.5-60 Hz. This categorization of the loads and household appliances is based on the contracts, consumers' desired level of comfort, and essential shedding levels. After dividing IR cost curve into *M* sections, at each section of cost cure, the market price can be calculated as follows where $C^{IR}$ ($/MW) is the cost offered by *GENCO* for IR. In each section of IR cost curve, the aggregator that offers the minimum price wins the competition.

$$price = \min\{C^{IR}, F_1^{DR}, F_2^{DR}, ..., F_G^{DR}\} \quad (4)$$

*3-Calculations of profit*

By competition between aggregators in different sections of IR cost cure, the aggregators and utility profits can be calculated as:

$$\text{Profit of } g^{th} \text{ aggregator} = \sum_{k=1}^{M} \Delta q_{k,g} . F_{k,g}^{DR} \quad (5)$$

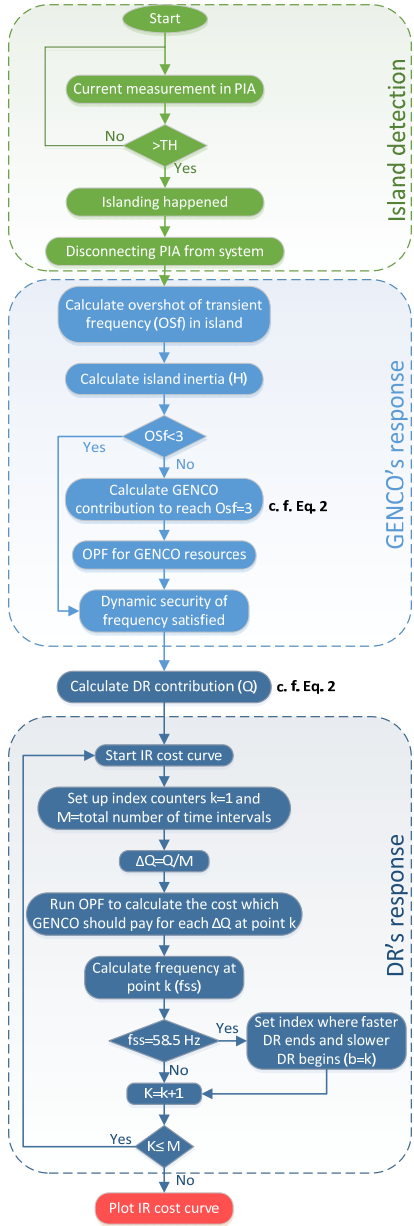

Figure 3. GENCO cost curve algorithm: determination of the CR cost curve for input into GENCO market algorithm.

$$\text{Total aggregators profit} = \sum_{k=1}^{M}\sum_{g=1}^{G}\Delta q_{k,g}.F_{k,g}^{DR} \quad (6)$$

$$\text{Utility's profit} = \sum_{k=1}^{M}\Delta q_k.C_k^{IR} - \sum_{k=1}^{M}\sum_{g=1}^{G}\Delta q_{k,g}.F_{k,g}^{DR} \quad (7)$$

$\Delta q_{k,g}$ is the shed load provided by the $g^{th}$ aggregator in the $k^{th}$ section ($k \in M$) of the CR cost curve partition (*MW*). $C_k^{IR}$ is the CR cost corresponding to section k.

## IV. TEST CASE AND SIMULATION RESULTS

As a case study, it has been supposed that the islanded microgird is similar to IEEE 24-bus model after losing the largest generation (350*MW* located in bus 23). This generator is the connection between microgrid and the rest of grid. In other words, the micrgord is connected to the rest of grid through bus 23. Technical data for generators is available in [10]. It is assumed that islanding detected using method in section II by detecting current in PIA. In the moment of islanding, the frequency sensors measure the frequency of 49 Hz. *GENCO*'s parameters for frequency regulations are: $\tau_g$ =0.2s, $\tau_T$ =0.5s, $D$ =0.8, $R$ =0.05 per unit in the island. Using equation (2) with the known amount of inertia for the island ($H$ =5s), *GENCO* and DRA's contribution are calculated to regulate the frequency. In the mentioned case study GENCO and DR's contribution are 282.9 and 67.1 *MW*, respectively, based on the algorithm shown in Figure 3. Figure 4 shows the transient frequency at the starting point of the IR cost curve before using the market for AS which the overshoot is less than 3 Hz. After the emergency IR (282.9 *MW*) had been supplied by GENCO to restore the island frequency above 57 Hz, algorithms of Figure 5 is performed to determine how DR could then be used to affect the remaining IR.

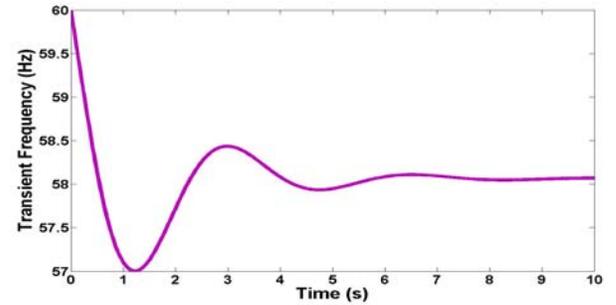

Figure 4. Transient frequency at the starting point of the CR cost curve

Figure 5 shows the cost of IR which GENCO should pay as an AS to regulate the island frequency. The red vertical line shows the border between the zones that should be supplied by faster and slower DR, based on the IEEE-1547 standard [2]. It means that 17.11 MW of the load should be shaded in faster DR zone less than 300s after the Islanding has been detected. Figure 6 demonstrates the bid prices for DR by two available aggregators in the island.

Figure 7 shows the aggregate quantity of the shed load by DRA 1 (blue color with shedding 38.08 *MW*) and DRA 2 (yellow color with shedding 29.02 *MW*) in the market. The profit of GENCO and DRAs are: $PR_{genco}^{tot}$ =1225$ $PR_{DRA,1}^{tot}$ =1372 $, and $PR_{DRA,2}^{tot}$ =1200$.

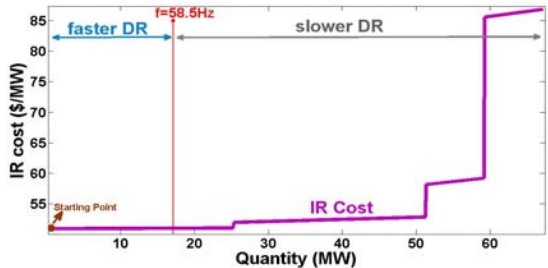
Figure 5. Cost of IR which should be paid by GENCO

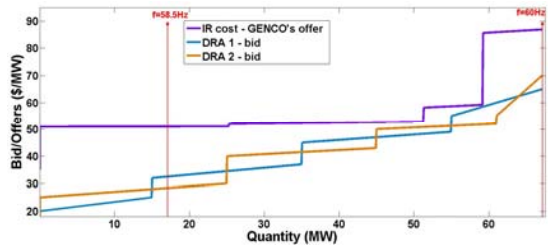
*Figure 6. Bid price for DR by two aggregators*

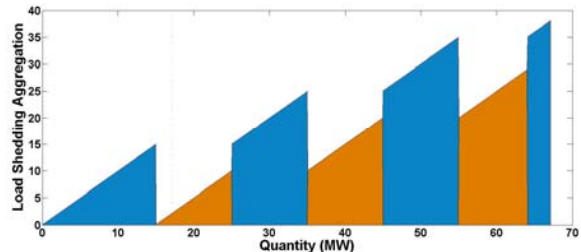
Figure 7. Aggregate quantity load shed by DRAs

## V. CONCLUSION

Considering the renewable energy escalation within the secondary network, a dynamic behavior of the grid needs to be considered with more precision. This paper present a new islanding detection method based on current detection in an added arm to the microgrid system which from one side eliminates unnecessary complexity of many other anti-islanding methods and also can overcome typical challenges associated with those methods. After islanding detection, instead of tripping the inverter-based units in the microgrid, this paper presents a market framework to provide AS in order to regulate the frequency in the islanded microgrid to run it as a standalone system. The simulation results show it clearly that the proposed method satisfy all the constraints for managing the frequency in the islanded situation.

References

[1] R. Teodorescu, M. Liserre, P. Rodríguez, Grid Converters for Photovoltaic and Wind Power Systems, John Wiley & Sons, 2010.
[2] IEEE 1547: IEEE standard for interconneting distributed resources with electric power systems, 2003.
[3] IEC 62116: Test procedure of islanding prevention measure for utilityinterconnected photovoltaic inverters, 2008.
[4] UL 1741: Inverters, Converters, Controllers and Interconnection System Equipment for Use With Distributed Energy Resources, 2005.
[5] S. Jang, K.H. Kim, "An islanding detection method for distributed generations using voltage unbalance and total harmonic distortion in current", IEEE Trans. Pow. Del., vol. 19, no. 2, pp. 745-752, Apr. 2004.
[6] M. Dietmannsberger; D. Schulz, "Impacts of Low Voltage Distribution Grid Codes on Ancillary Services and Anti-Islanding-Detection of Inverter-Based Generation," in IEEE Transactions on Energy Conversion , vol.PP, no.99, pp.1-1
[7] A. Zeinalzadeh, R. Ghorbani, J. Yee, Voltage reliability in the presence of high penetration of photovoltaic systems, IEEE American Control Conference (ACC), Boston, USA, 2016.
[8] ]Y. J. Kim, E. Fuentes and L. K. Norford, "Experimental Study of Grid Frequency Regulation Ancillary Service of a Variable Speed Heat Pump," in IEEE Transactions on Power Systems, vol. 31, no. 4, pp. 3090-3099, July 2016.
[9] S. Wang, L. Han, D. Wang, M. Shahidehpour and Z. Li, "Hierarchical charging management strategy of plug-in Hybrid Electric Vehicles to provide regulation service," 2012 3rd IEEE PES Innovative Smart Grid Technologies Europe (ISGT Europe), Berlin, 2012, pp. 1-6.
[10] M. Motalleb, M. Thornton, E. Reihani and R. Ghorbani, "A nascent market for contingency reserve services using demand response," Applied Energy, pp. 985-995, 2016.
[11] M. Motalleb, M. Thornton, E. Reihani and R. Ghorbani, "Providing frequency regulation reserve services using demand response scheduling," Energy Conversion and Management, vol. 124, pp. 439-452, 2016.
[12] E. Reihani, M. Motalleb, M. Thornton and R. Ghorbani, "A novel approach using flexible scheduling and aggregation to optimize demand response in the developing interactive grid market architecture," Applied Energy, 183, 445-455., vol. 183, pp. 445-455, 2016.
[13] P. Babahajiani; Q. Shafiee and Hassan Bevrani, "Intelligent Demand Response Contribution in Frequency Control of Multi-area Power Systems," IEEE Trans. Smart Grid, June 2016.
[14] S. A. Pourmousavi and M. H. Nehrir, "Real-Time Central Demand Response for Primary Frequency Regulation in Microgrids," IEEE Trans. Smart Grid., vol. 3, pp. 1988–1996, Dec 2012.
[15] I. Beil, I. Hiskens and S. Backhaus, "Frequency Regulation From Commercial Building HVAC Demand Response," Proceedings of the IEEE, Vol. 104, No. 4, April 2016.
[16] A. Eshraghi and R. Ghorbani, "Islanding detection and transient over voltage mitigation using wireless sensor networks," IEEE Power & Energy Society General Meeting, July 2015.
[17] A. Eshraghi and R. Ghorbani, "Islanding detection and over voltage mitigation using controllable loads," Sustainable Energy, Grids and Networks, vol. 6, pp. 425-135, June 2016.
[18] D. Kirschen, and G. Strbac, Fundamentals of Power System Economics, John Wiley & Sons Ltd, 2004.

[19] A. Eshraghi, B. Maham, Z. Han, M. Banagar, Efficiency and coverage improvement of active RFID two-hop relay systems, in: IEEE Wireless Communications and Networking Conference, WCNC'14, Istanbul, Turkey, 2014, pp. 2002–2007.
[20] A. Zeinalzadeh and Vijay Gupta, Minimizing risk of load shedding and renewable energy curtailment in a microgrid with energy storage, ArXiv e-prints, arXiv: 1611.08000, Nov. 2016.
[21] A. Zeinalzadeh, and Vijay Gupta, Pricing energy in the presence of renewables, ArXiv e-prints, arXiv:1611.08006, Nov 2016.
[22] A. Zeinalzadeh, R. Ghorbani, E. Reihani, Optimal power flow problem with energy storage voltage and reactive power control, The 45th ISCIE International Symposium on Stochastic Systems Theory and Its Applications, Okinawa, 2013.

.